\documentstyle[aps,preprint,graphicx]{revtex}
\tighten
\draft

\begin{document}

\def\simge{\hspace*{0.2em}\raisebox{0.5ex}{$>$}
     \hspace{-0.8em}\raisebox{-0.3em}{$\sim$}\hspace*{0.2em}}
\def\simle{\hspace*{0.2em}\raisebox{0.5ex}{$<$}
     \hspace{-0.8em}\raisebox{-0.3em}{$\sim$}\hspace*{0.2em}}
\def\bra#1{{\langle#1\vert}}
\def\ket#1{{\vert#1\rangle}}
\def\coeff#1#2{{\scriptstyle{#1\over #2}}}
\def\undertext#1{{$\underline{\hbox{#1}}$}}
\def\hcal#1{{\hbox{\cal #1}}}
\def\sst#1{{\scriptscriptstyle #1}}
\def\eexp#1{{\hbox{e}^{#1}}}
\def\rbra#1{{\langle #1 \vert\!\vert}}
\def\rket#1{{\vert\!\vert #1\rangle}}
\def\lsim{{ <\atop\sim}}
\def\gsim{{ >\atop\sim}}
\def\nubar{{\bar\nu}}
\def\psibar{{\bar\psi}}
\def\Gmu{{G_\mu}}
\def\alr{{A_\sst{LR}}}
\def\wpv{{W^\sst{PV}}}
\def\evec{{\vec e}}
\def\notq{{\not\! q}}
\def\notk{{\not\! k}}
\def\notp{{\not\! p}}
\def\notpp{{\not\! p'}}
\def\notder{{\not\! \partial}}
\def\notcder{{\not\!\! D}}
\def\notA{{\not\!\! A}}
\def\notv{{\not\!\! v}}
\def\Jem{{J_\mu^{em}}}
\def\Jana{{J_{\mu 5}^{anapole}}}
\def\nue{{\nu_e}}
\def\mn{{m_\sst{N}}}
\def\mns{{m^2_\sst{N}}}
\def\me{{m_e}}
\def\mes{{m^2_e}}
\def\mq{{m_q}}
\def\mqs{{m_q^2}}
\def\mz{{M_\sst{Z}}}
\def\mzs{{M^2_\sst{Z}}}
\def\ubar{{\bar u}}
\def\dbar{{\bar d}}
\def\sbar{{\bar s}}
\def\qbar{{\bar q}}
\def\sstw{{\sin^2\theta_\sst{W}}}
\def\gv{{g_\sst{V}}}
\def\ga{{g_\sst{A}}}
\def\pv{{\vec p}}
\def\pvs{{{\vec p}^{\>2}}}
\def\ppv{{{\vec p}^{\>\prime}}}
\def\ppvs{{{\vec p}^{\>\prime\>2}}}
\def\qv{{\vec q}}
\def\qvs{{{\vec q}^{\>2}}}
\def\xv{{\vec x}}
\def\xpv{{{\vec x}^{\>\prime}}}
\def\yv{{\vec y}}
\def\tauv{{\vec\tau}}
\def\sigv{{\vec\sigma}}
\def\sst#1{{\scriptscriptstyle #1}}
\def\gpnn{{g_{\sst{NN}\pi}}}
\def\grnn{{g_{\sst{NN}\rho}}}
\def\gnnm{{g_\sst{NNM}}}
\def\hnnm{{h_\sst{NNM}}}

\def\xivz{{\xi_\sst{V}^{(0)}}}
\def\xivt{{\xi_\sst{V}^{(3)}}}
\def\xive{{\xi_\sst{V}^{(8)}}}
\def\xiaz{{\xi_\sst{A}^{(0)}}}
\def\xiat{{\xi_\sst{A}^{(3)}}}
\def\xiae{{\xi_\sst{A}^{(8)}}}
\def\xivtez{{\xi_\sst{V}^{T=0}}}
\def\xivteo{{\xi_\sst{V}^{T=1}}}
\def\xiatez{{\xi_\sst{A}^{T=0}}}
\def\xiateo{{\xi_\sst{A}^{T=1}}}
\def\xiva{{\xi_\sst{V,A}}}

\def\rvz{{R_\sst{V}^{(0)}}}
\def\rvt{{R_\sst{V}^{(3)}}}
\def\rve{{R_\sst{V}^{(8)}}}
\def\raz{{R_\sst{A}^{(0)}}}
\def\rat{{R_\sst{A}^{(3)}}}
\def\rae{{R_\sst{A}^{(8)}}}
\def\rvtez{{R_\sst{V}^{T=0}}}
\def\rvteo{{R_\sst{V}^{T=1}}}
\def\ratez{{R_\sst{A}^{T=0}}}
\def\rateo{{R_\sst{A}^{T=1}}}

\def\mro{{m_\rho}}
\def\mks{{m_\sst{K}^2}}
\def\mpi{{m_\pi}}
\def\mpis{{m_\pi^2}}
\def\mom{{m_\omega}}
\def\mphi{{m_\phi}}
\def\Qhat{{\hat Q}}

\def\FOS{{F_1^{(s)}}}
\def\FTS{{F_2^{(s)}}}
\def\GAS{{G_\sst{A}^{(s)}}}
\def\GES{{G_\sst{E}^{(s)}}}
\def\GMS{{G_\sst{M}^{(s)}}}
\def\GATEZ{{G_\sst{A}^{\sst{T}=0}}}
\def\GATEO{{G_\sst{A}^{\sst{T}=1}}}
\def\mdax{{M_\sst{A}}}
\def\mustr{{\mu_s}}
\def\rsstr{{r^2_s}}
\def\rhostr{{\rho_s}}
\def\GEG{{G_\sst{E}^\gamma}}
\def\GEZ{{G_\sst{E}^\sst{Z}}}
\def\GMG{{G_\sst{M}^\gamma}}
\def\GMZ{{G_\sst{M}^\sst{Z}}}
\def\GEn{{G_\sst{E}^n}}
\def\GEp{{G_\sst{E}^p}}
\def\GMn{{G_\sst{M}^n}}
\def\GMp{{G_\sst{M}^p}}
\def\GAp{{G_\sst{A}^p}}
\def\GAn{{G_\sst{A}^n}}
\def\GA{{G_\sst{A}}}
\def\GETEZ{{G_\sst{E}^{\sst{T}=0}}}
\def\GETEO{{G_\sst{E}^{\sst{T}=1}}}
\def\GMTEZ{{G_\sst{M}^{\sst{T}=0}}}
\def\GMTEO{{G_\sst{M}^{\sst{T}=1}}}
\def\lamd{{\lambda_\sst{D}^\sst{V}}}
\def\lamn{{\lambda_n}}
\def\lams{{\lambda_\sst{E}^{(s)}}}
\def\bvz{{\beta_\sst{V}^0}}
\def\bvo{{\beta_\sst{V}^1}}
\def\Gdip{{G_\sst{D}^\sst{V}}}
\def\GdipA{{G_\sst{D}^\sst{A}}}
\def\fks{{F_\sst{K}^{(s)}}}
\def\FIS{{F_i^{(s)}}}
\def\fpi{{F_\pi}}
\def\fk{{F_\sst{K}}}

\def\RAp{{R_\sst{A}^p}}
\def\RAn{{R_\sst{A}^n}}
\def\RVp{{R_\sst{V}^p}}
\def\RVn{{R_\sst{V}^n}}
\def\rva{{R_\sst{V,A}}}
\def\xbb{{x_B}}

\def\PR#1{{{\em   Phys. Rev.} {\bf #1} }}
\def\PRC#1{{{\em   Phys. Rev.} {\bf C#1} }}
\def\PRD#1{{{\em   Phys. Rev.} {\bf D#1} }}
\def\PRL#1{{{\em   Phys. Rev. Lett.} {\bf #1} }}
\def\NPA#1{{{\em   Nucl. Phys.} {\bf A#1} }}
\def\NPB#1{{{\em   Nucl. Phys.} {\bf B#1} }}
\def\AoP#1{{{\em   Ann. of Phys.} {\bf #1} }}
\def\PRp#1{{{\em   Phys. Reports} {\bf #1} }}
\def\PLB#1{{{\em   Phys. Lett.} {\bf B#1} }}
\def\ZPA#1{{{\em   Z. f\"ur Phys.} {\bf A#1} }}
\def\ZPC#1{{{\em   Z. f\"ur Phys.} {\bf C#1} }}
\def\etal{{{\em   et al.}}}

\def\delalr{{{delta\alr\over\alr}}}
\def\pbar{{\bar{p}}}
\def\lamchi{{\Lambda_\chi}}
\newcommand{\amulbl}{a_\mu^{\sst{LL}}({\rm had})}
\newcommand{\amulbllo}{a_\mu^{\sst{LL}}({\rm had})_{\rm l.o.}}

\preprint{
\noindent
\hfill
\begin{minipage}[t]{3in}
\begin{flushright}
CALT-68-2372\\
MAP-282\\
hep-ph/0201297\\
\vspace*{.7in}
\end{flushright}
\end{minipage}
}

\title{Hadronic Light-by-Light Contribution to Muon $g-2$ in Chiral
Perturbation Theory}

\author{M.J. Ramsey-Musolf$^{a,b}$  and
Mark B. Wise$^a$
\\[0.3cm]
}
\address{
$^a$ California Institute of Technology,
Pasadena, CA 91125\ USA\\
$^b$ Department of Physics, University of Connecticut, Storrs, CT 06269\ USA
}


\maketitle

\begin{abstract}

We compute the hadronic light-by-light scattering contributions to the muon
anomalous magnetic moment, $\amulbl$, in chiral perturbation theory that
are enhanced by large
logarithms and a factor of $N_C$. They depend on a low-energy constant entering
pseudoscalar meson decay
into a charged lepton pair. The uncertainty introduced by this constant is
$\pm 60\times 10^{-11}$, which is comparable in
magnitude to the present uncertainty entering the leading-order vacuum
polarization
contributions to
the anomalous moment. It may be reduced to some extent through an improved
measurement of the
$\pi^0\to e^+ e^-$ branching ratio. However, the dependence of $\amulbl$ on
non-logarithmically enhanced effects cannot be constrained except through
the measurement of
the anomalous moment itself. The extraction of information on new physics
would require  a
future experimental value for the anomalous moment differing significantly
from the 2001 result reported by the E821 collaboration.

\end{abstract}



\vspace{0.3cm}

\pagenumbering{arabic}

The recently reported measurement of the muon anomalous magnetic moment
$a_\mu$ by the E821
collaboration\cite{brown} has generated considerable excitement about
possible evidence for new
physics. The interpretation of the result, however, depends in part on a
reliable treatment of
hadronic  contributions to $a_\mu$ which arise at two- and three-loop
order. While the vacuum polarization contribution can be constrained by
$e^+e^-$ experiments and $\tau$ decays and appears to be under adequate
theoretical
control\cite{narrison}, recent analysis\cite{knecht,deraf} of the hadronic
light-by-light
scattering contribution
$\amulbl$ have uncovered a sign error in
previous calculations \cite{hay95,hay96,hay98,bij95,bij96} of the dominant,
pseudoscalar pole
term. The resulting
sign change  reduces the $2.6\sigma$ deviation of $a_\mu$ from the Standard
Model prediction reported
in \cite{brown} by one standard deviation, thereby modifying considerably
the original interpretation of the result\cite{hay01,bij01}.

The commonly quoted values for $\amulbl$ (after incorporating the corrected
overall sign) rely
on model treatments of the off shell $\pi^0\gamma^\ast\gamma^\ast$,
$\eta\gamma^\ast\gamma^\ast$ and $\eta^{\prime}\gamma^\ast\gamma^\ast$
interactions. While the amplitude for pseudscalar decay into two real
photons is dictated by the
chiral anomaly, the off-shell amplitudes relevant for $\amulbl$ are
affected by non-perturbative
strong interactions whose effects cannot yet be computed with sufficient
precision from first
principles in QCD. Similarly, the contributions from other hadronic
intermediate states
besides the $\pi^0$, $\eta$, and $\eta^{\prime}$ cannot be computed
reliably at present.
The analysis of $\amulbl$ falls naturally under the purview of chiral
perturbation theory ($\chi$PT), which provides systematic,
model-independent framework for
parameterizing presently incalculable hadronic effects. The static quantity
$\amulbl$ has an expansion
in powers of $p/\Lambda$, where $p$ is a small  mass of order $m_\mu$ or
$m_\pi$ and $\Lambda$ is a
hadronic scale, typically taken to be $\sim 4\pi F_\pi\sim
1$GeV\footnote{We treat $m_{\mu}$ and $m_{\pi}$ to be of the same order and
take both to be small compared
with $\Lambda$.}. The
coefficients appearing in the
expansion depend  in part on {\em a priori} unknown \lq\lq low-energy
constants" (LEC's), which
parameterize the effects of non-perturbative short distance physics. In
principle, the LEC's may be
determined from an appropriate set of experimental measurements.

In this Letter, we perform a $\chi$PT calculation of $\amulbl$ including
all the
logarithmically enhanced contributions to $\amulbl$
that arise at order $N_C \alpha^3
p^2/\Lambda^2$, where $N_C$ is the number of quark colors. We identify the
dependence of
$\amulbl$ on the large logarithms of
$\Lambda /p$ as well as on the relevant LEC's.  The result for the large
$\ln^2$ term -- which
is determined entirely by gauge-invariance and the chiral anomaly -- was
first given in Ref.
\cite{deraf} and was used to
uncover the sign error in previous model calculations.  However, only part
of the large $\ln$ term is fixed
by symmetry considerations. It receives an additional contribution
involving $\chi$, a LEC entering the
rate for pseudoscalar decay into leptons. The $\chi$-dependent piece was
also computed in Ref.
\cite{deraf}. An analytic calculation of the $\chi$-independent large $\ln$
term was
performed by the authors of Ref.
\cite{blokland}, who used a model for the off-shell
$P\gamma^\ast\gamma^\ast$ form factor to regulate
the two-loop amplitude and assumed $m_\mu$ was almost equal to $m_\pi$. The
results of the
latter calculation also contain
a model prediction for the non-logarithmic ${\cal O}(N_C\alpha^3
p^2/\Lambda^2)$ contribution to $\amulbl$.

In what follows, we carry out a consistent $\chi$PT treatment of $\amulbl$,
providing the first
complete, model-independent analysis of the terms enhanced by large
logarithms through ${\cal
O}(N_C\alpha^3 p^2/\Lambda^2)$. The sum of these terms is known, since
existing measurements for
$\eta\to\mu^+\mu^-$ and $\pi^0\to e^+ e^-$ branching ratios fix the value
of $\chi$. However, the
uncertainty in $\amulbl$ from the error in $\chi$ is significant:
$\pm 60\times 10^{-11}$, which is
roughly the same size as the present
theoretical uncertainty in the leading-order vacuum polarization
contributions to $a_\mu$. In principle,
an improved measurement of the
$\pi^0\to e^+ e^-$ branching ratio could reduce this source of uncertainty.

A more serious consideration
involves the contributions to $\amulbl$ that are beyond the order at which
we compute.
These include effects of order ${\cal O}(\alpha^3 p^2/\Lambda^2)$ which are
not enhanced by a factor
of $N_C$ and order ${\cal O}(N_C\alpha^3 p^2/\Lambda^2)$ terms that are not
enhanced by large logarithms.
We parameterize all these effects in
terms of ${\tilde C}$. For simplicity we will
refer to $\tilde C$ as a low energy constant even though it has nonanalytic
dependence on the quark masses.
At present, ${\tilde C}$ cannot be determined in a model-independent way
without reliance
on the measurement of $a_\mu$ itself. Even the sign of the
${\tilde C}$-dependent term cannot presently be fixed in an
model-independent manner.
The presence of this
LEC renders the interpretation of $a_\mu$ in terms of new physics
problematic, since the size of the ${\tilde C}$-dependent contribution could be
as large as the present experimental error in $a_\mu$\cite{brown}. Below we
discuss
the conditions under which one might still be able to extract information
on new physics from an $a_\mu$
measurement.

To leading order in $N_C$, the lowest-order chiral
contributions to $\amulbl$,  enhanced by large logarithms, arise from the
two- and one-loop graphs of
Fig 1. Taking $m_\pi$ and $m_\mu$  as being of ${\cal O}(p)$, the leading,
large logarithmic contributions
arise at order
$N_C\alpha^3 p^2/\Lambda^2$. The two-loop graphs (Fig. 1a) contain an
overall, superficial cubic
divergence as well as a linearly-divergent one-loop subgraph involving two
photons and a muon line. The
latter must be regulated by adding the appropriate one-loop counterterm
(ct) (Fig. 1b). The one-loop graphs also
contain an insertion of $\chi(\mu)$. The sum of these graphs contains a
residual divergence, which must
be removed by the appropriate magnetic moment ct (Fig. 1c). Associated with
this ct is a
finite piece which, as discussed above, can only be fixed in a
model-independent way by
the measurement of $a_\mu$ itself. Additional contributions also arise from
the graphs such as those
appearing in Fig. 2. Although subdominant in $N_C$ counting the
contribution of the charged pion loop
in 2b (and related diagrams) is leading order in $p /\Lambda$. The order
${\cal O}(\alpha^3 N_C^0)$ term arising from the three loop graphs with a
charged pion loop -- which
we denote by $\amulbllo$ --
has been computed in Ref. \cite{kin85,hay95}. The result is finite and
contains no large logarithms. We will include this contribution when we
compare
the theoretical prediction for $\amulbl$ with experiment.

At order $\alpha^3 p^2/\Lambda^2$ there will be contributions to $\amulbl$
from higher dimension
operators inserted at the vertices in the charged pion loop graph of Fig.2
and from four-loop graphs containing an
additional hadronic loop. Some of these should contain large logarithms.
However, these are suppressed by a factor
of $N_c$ compared to the logarithmically enhanced pieces that we compute.
As remarked earlier we absorb these and
many other subdominant contributions ({\em e.g.}, Fig. 2a) into $\tilde C$
and do not discuss them further here.

As inputs for amplitudes of Fig. 1, we require the Wess-Zumino-Witten
$P\gamma\gamma$
interaction Lagrangian\cite{wzw}:
\begin{equation}
\label{eq:wzw}
{\cal L}_{WZW} = {\alpha N_C\over 24\pi
F_\pi}\epsilon_{\mu\nu\alpha\beta}F^{\mu\nu}
F^{\alpha\beta}\left(\pi^0+\frac{1}{\sqrt{3}}\eta\right) +\cdots
\end{equation}
as well as the leading-order operator contributing to the decays
$P\to\ell^+\ell^-$\cite{savage}:
\begin{eqnarray}
\label{eq:wise}
{\cal L}_{P\ell^+\ell^-} &= &{i3N_C\alpha^2\over
96\pi^2}{\bar\mu}\gamma^\lambda\gamma_5\mu\Bigl[
\chi_1{\mbox{Tr}}\Bigl(Q^2\Sigma^\dag D_\lambda\Sigma-Q^2
D_\lambda\Sigma^\dag\Sigma\Bigr)\\
&&
\nonumber
+\chi_2{\mbox{Tr}}\Bigl(Q\Sigma^\dag QD_\lambda\Sigma-QD_\lambda\Sigma^\dag
Q\Sigma\Bigr)\Bigr]\\
\label{eq:wise2}
&=&-{N_C\alpha^2\over 48\pi^2
F_\pi}(\chi_1+\chi_2){\bar\mu}\gamma^\lambda\gamma_5\mu
\left(\partial_\lambda
\pi^0+\frac{1}{\sqrt{3}}\partial_\lambda\eta\right)+\cdots\ \ \ .
\end{eqnarray}
where
\begin{equation}
\Sigma=\exp\left(\sum\lambda^a\pi^a/F_\pi\right)
\end{equation}
Here, $F_\pi=92.4$ MeV is the pion decay constant, $\lambda^a$ denote the
Gell-Mann SU(3) matrices, and
$D_\lambda$ is the covariant derivative. Note that in contrast to the
conventions of
Ref. \cite{savage}, we have made the $N_C$-dependence of the LEC's $\chi_i$
explicit
for the sake of clarity.

In computing the loop amplitudes involving these operators, it is important
to employ a regulator which
maintains the consistent power-counting of the chiral expansion. To that
end, we employ dimensional
regularization, where we continue only momenta (and not Dirac matrices)
into $d=4-2\epsilon$ dimensions. The relation between bare and renormalized
couplings is,
\begin{equation}
\label{eq:oneloopct}
\chi_1^0+\chi_2^0 = \chi_1+\chi_2 - {6 \over \epsilon} \equiv
\chi(\mu)-{6 \over \epsilon}\ \ \ .
\end{equation}
Using Eq.
(\ref{eq:oneloopct}) and
adding the amplitudes for Fig. 1a,b, we obtain the divergent part of the
two-loop amplitude
\begin{equation}
\label{eq:twoloop}
{\cal M} = -e\left({\alpha\over\pi}\right)^3\left({N_C\over
3}\right)^2\left({m_\mu\over
F_\pi}\right)^2\left({1\over
32\pi^2}\right)
\left[{3\over 2\epsilon^2}
-\left(\frac{\chi(\mu)}{2}+\frac{3}{4}\right)\frac{1}{\epsilon}\right]{\bar
u} {i\sigma_{\alpha\beta}
q^\alpha\varepsilon^\beta \over 2 m_{\mu}}u
\ \ \ ,
\end{equation}
where $q^{\alpha}$ and $\varepsilon^\beta$ are the photon momentum and
polarization, respectively.
We remove this divergence using a magnetic moment counterterm. The bare
coupling $C_0$ and renormalized coupling $C(\mu)$ are related by
\begin{eqnarray}
\label{eq:mren}
{\cal M}_0 & = & e\left({\alpha\over\pi}\right)^3 \left({N_C\over
3}\right)^2\left({m_\mu\over
F_\pi}\right)^2\left({1\over
32\pi^2}\right) C_0 \bar u {i\sigma_{\alpha\beta} q^\alpha\varepsilon^\beta
\over 2 m_{\mu}}u\\
\label{eq:c0def}
C_0 & = & C(\mu) + \left[{3 \over 2\epsilon^2}
-\left(\frac{\chi(\mu)}{2}+\frac{3}{4}\right)\frac{1}{\epsilon}\right]\ \ \ .
\end{eqnarray}

The light-by-light contribution to the anomalous magnetic moment $\amulbl$
is a
physical quantity and has no dependence on the subtraction point $\mu$. The
$\mu$-dependence of the diagrams
cancels that of the couplings $C(\mu)$ and $\chi(\mu)$. To obtain the
$\mu$-dependence of the couplings we
require that the bare Green's functions corresponding to the sum of Figs.
1a-c and the to the
$P\ell^+\ell^-$ one-loop subgraphs,
respectively, be independent of the subtraction scale. Doing so leads to a
coupled set of renormalization group equations for $\chi(\mu)$ and $C(\mu)$:
\begin{eqnarray}
\label{eq:rg}
\mu\frac{d\chi}{d\mu} & = & -12 \\
\mu\frac{d C}{d\mu} & = & -3 -\chi \ \ \ .
\end{eqnarray}
The solution is
\begin{eqnarray}
\label{eq:solution}
\chi(\mu) &=& 12\ln(\mu_0/\mu) +\chi(\mu_0)\\
C(\mu) &=& 6\ln^2(\mu_0/\mu)+[\chi(\mu_0)+3]\ln(\mu_0/\mu) + C(\mu_0)\ \ \ .
\end{eqnarray}

At a scale $\mu_0=\Lambda\sim 1$ GeV, the constants $C(\mu_0)$ and
$\chi(\mu_0)$ contain
no large logarithms of
the form $\ln^k(\Lambda /p)$ ($k=1,2$) where $p$ is around $m_{\mu}$ or
$m_\pi$.
For $\mu$ of ${\cal O}(p)$,
however, the Feynman diagrams contain no such large logarithms, and they
live entirely in $C(\mu)$ and
$\chi(\mu)
$. Hence, the resulting expression
for $\amulbl$ is, in the $\overline{MS}$ scheme,
\begin{eqnarray}
\nonumber
\amulbl &=& \amulbllo\\
\label{eq:main}
&+& {3\over 16} \left({\alpha\over\pi}\right)^3\left({m_\mu\over
F_\pi}\right)^2\left({N_C\over 3\pi}
\right)^2\Bigl\{\ln^2\left(\frac{\Lambda}{\mu}\right)
+\left[-f(r)+\frac{1}{2}+\frac{1}{6}\chi(\Lambda)
\right]\ln\left(\frac{\Lambda}{\mu}\right)+{\tilde C}\Bigr\} ,
\end{eqnarray}
where $\mu$ is of order $p$ and
could be set equal to either
$m_\mu$ or $m_\pi$. Recall that
$\Lambda\sim 4\pi F_\pi\sim 1$ GeV. The function
$f(r)$, with $r=m_\pi^2/m_\mu^2$, arises from the one loop diagram with a
coupling proportional to
$\chi({\mu})$ (Fig. 1b) and is given by\footnote{Since we only compute the
terms enhanced by large
logarithms to get $f$ we replace $\chi(\mu)$ by
$12{\rm ln}(\Lambda/\mu)$.}
\begin{equation}
\label{eq:fr}
f(r)=\ln\left(\frac{m_\mu^2}{\mu^2}\right)+\frac{1}{6}r^2\ln r
-\frac{1}{6}(2r+13)\\
+\frac{1}{3}(2+r)\sqrt{r(4-r)}\cos^{-1}\left(\frac{\sqrt{r}}{2}\right)\ \ \ .
\end{equation}
Note that we have absorbed all the remaining terms, including those
proportional to $C(\Lambda)$ and
$\chi({\Lambda})$ not enhanced by large logarithms, into ${\tilde C}$.

The logarithmically enhanced hadronic light-by-light contributions to
$a_\mu$ are renormalization scheme-independent. However, the values of
$\chi(\Lambda)$, $f(r)$, and the constant appearing in the renormalization
group equation for $C$ (leading to the factor of $1/2$ in the $\ln$ term in
Eq. (\ref{eq:main}) ) depend on one's choice of scheme\footnote{We will
discuss
the issue of scheme dependence and give more
details of the calculation in a future publication.}. This scheme-dependence
cancels in the sum of their contributions to $\amulbl$. For our calculations,
we adopted a scheme in which the loop integrals were evaluated in
$d$-dimensions with $d>4$ (corresponding to $\epsilon < 0$), while the Dirac
matrices and photon polarization indices were taken as four-dimensional. For
this choice we have $\eta_{\mu\nu}{\rm Tr}(\gamma^\mu\gamma^\nu)=16$ instead
of $4d$. Moreover, the value of $\chi(\Lambda)$ in this
scheme is the same as that in \cite{savage} but is four less than the
$\chi(\Lambda)$ used in \cite{ame01}. An alternative, equally convenient
scheme is
again to treat the Dirac matrices, epsilon tensors, and photon polarizations as
four dimensional; take $d<4$ ($\epsilon> 0$); and rewrite (\ref{eq:wise2}) as
\begin{equation}
i{N_C\alpha^2\over 48 \pi^2 F_\pi}{(\chi_1+\chi_2)\over
6}\epsilon^{\mu\alpha\nu\lambda}
{\bar\mu} \gamma_\mu\gamma_\beta\gamma_\mu \mu\
\eta^\beta_\alpha\partial_\lambda
\left(\pi^0+\frac{1}{\sqrt{3}}\eta\right)+\cdots\ \ \ ,
\end{equation}
where the metric tensor $\eta_{\alpha\beta}$ is $d$-dimensional. In this scheme
the value of $\chi(\Lambda)$ is still the same as in \cite{savage}, but
$f(r) \rightarrow f(r)+3/2$ and the $-3$ in the renormalization group
equation for $C$
becomes $-12$. Notice that the total value for the logarithmically enhanced
contribution to
$\amulbl$ is unchanged.

As a check on the result in Eq. (\ref{eq:main}), one may compute the one-
and two-loop amplitudes
with the insertion of $\chi(\Lambda)$ in Fig. 1b and $C(\Lambda)$ in Fig.
1c. In this case, all of
the large logarithms arise from the Feynman amplitudes and not from the
operator coefficients. Using
an explicit calculation, we have verified in the limit $m_\pi\to 0$ that
this procedure exactly reproduces
the expression in Eq. (\ref{eq:main}). We note that the $\ln^2$ term and
the term proportional to
$\chi$ agree with the expression in Ref. \cite{deraf}.

Chiral perturbation theory can be used for the
$\eta \rightarrow \mu^+ \mu^-$ amplitude \cite{savage}, and the  LEC
$\chi(\Lambda)$ can be deduced from
the measured $\eta\to\mu^+\mu^-$ branching
ratio \cite{pdg}. This yields \cite{ame01} $\chi(1 {\rm GeV}) =
-14^{+4}_{-5}$ or $-39^{+5}_{-4}$,
where we have scaled the results in Ref. \cite{ame01} from $\Lambda=m_\rho$
up to $\Lambda =1$ GeV and subtracted four.
Note that because the $\eta\to\mu^+\mu^-$ branching ratio is a quadratic
function of $\chi$, two different
values for this LEC may be extracted from experiment.

Our calculation of the logarithmically enhanced contributions to $\amulbl$
only involved the use of chiral
$SU(2)_L \times SU(2)_R$ while this extraction of the LEC involves the use
of chiral $SU(3)_L
\times SU(3)_R$. Since one expects chiral perturbation theory to work
better in the case where only the
pions are treated as light it is desirable to have an extraction of $\chi$
that only relies on chiral
$SU(2)_L \times SU(2)_R$. This can be done using the measured $\pi^0\to e^+
e^-$ branching ratio which
yields \cite{ame01} $ \chi(1 {\rm GeV}) =  -29^{+25}_{-16}$ or
$+74^{+16}_{-25}$.
Unfortunately, the errors on the extracted $\chi(1 {\rm GeV})$ are very
large in this case.
A more precise determination of the $\pi^0\to e^+ e^-$ branching ratio
could reduce the theoretical
uncertainty in $\chi(1{\rm GeV})$.

Model calculations for $\amulbl$ differ from our analysis
typically through insertion of form factors at the
$P\gamma^\ast\gamma^\ast$ vertices
obtained from the WZW interaction in Eq. (\ref{eq:wzw}). For example,
one widely-followed model employs
form factors based on a vector meson dominance
picture. This approach -- known as resonance saturation -- may also be
used to obtain $\chi$, giving
\cite{ame01} $\chi(1 {\rm GeV})_{\mbox{res sat}} \simeq -17 $ .
In a similar vein, one may analyze this LEC at leading order in $N_C$, where
it depends on a sum over an infinite tower of vector meson
resonances\cite{knecht2}.
Using a model-dependent form factor for the sum over vector
resonances that at short distances is consistent with the properties of QCD
gives
$\chi(1 {\rm GeV})_{N_C,\ \mbox{res sat}}=-16\pm 5$.
This result provides some quantitative support for phenomenological models
since it is close
to the value  $\chi(1 {\rm GeV}) \simeq -14$ obtained
from  experiment. However, one would not want to draw conclusions about the
validity of the Standard Model using such a model-dependent approach.

Using $\chi(1 {\rm GeV}) = -14^{+4}_{-5}$ as input, setting
$\mu=m_\mu$,  and adding the large $\ln^2$ and $\ln$ terms in Eq.
(\ref{eq:main}), we obtain
\begin{equation}
\label{eq:nonan}
\amulbl_{\rm log} =   \left(57^{+50}_{-60}\right)\times 10^{-11}\ \ \ .
\end{equation}
We emphasize that inclusion of {\em both} the $-f(r)+1/2$ and $\chi/6$
parts of the $\ln(\Lambda/\mu)$
term is crucial to obtaining an accurate numerical result for $\amulbl_{\rm
log}$. If, for example,
one were to keep only the dependence on $\chi(\Lambda)$, one would find
substantial cancelations between
the $\ln^2$ and $\ln$ contributions. The presence of the calculable
$-f(r)+1/2$ term, however,
substantially mitigates these cancelations.

We observe that the central value in Eq. (\ref{eq:nonan})  is roughly a
factor of two larger than obtained
in model calculations for the $\pi^0$ contribution, and that the uncertainty
is about a third the size of the present experimental error in
$a_\mu$\cite{brown}.
After the full E821 data set is
analyzed, the uncertainty in Eq. (\ref{eq:nonan}) will be comparable to the
anticipated experimental error. As noted
above, improved measurements of the $\pi^0\to e^+ e^-$ branching ratio
could reduce the theoretical
uncertainty in the large logarithmic contributions to $\amulbl$. Using the
other value of $\chi$
obtained from the $\eta\to\mu^+\mu^-$ branching ratio, $\chi(1 {\rm GeV}) =
-39^{+5}_{-4}$, leads
to $ \amulbl_{\rm log} = \left(-190^{+60}_{-50}\right)\times 10^{-11}\ \ \ $.
Although there exists a strong theoretical prejudice in favor of the first
solution [Eq. (\ref{eq:nonan})]
based on both the resonance saturation model for $\chi$ and $\amulbl$ as
well as consistency between the
values of $\chi$ obtained from  the $\eta\to\mu^+\mu^-$ and $\pi^0\to e^+
e^-$ branching ratios, one cannot
rule out the second value for $\amulbl_{\rm log}$.

Adding in the $O(N_C^0 \alpha^3)$ charged pion loop contribution
\cite{hay95,kin85},
$\amulbllo=-44.6 \times 10^{-11}$, to that in Eq. (\ref{eq:nonan}) gives
the following
$\chi$PT  expression for $\amulbl$,
\begin{equation}
\label{eq:chiral}
\amulbl =   \left(13^{+50}_{-60}\ +31{\tilde C}\right)\times 10^{-11}\ \ \ .
\end{equation}

The largest uncertainty in the expression for $\amulbl$ above
arises from the subdominant terms that have not been calculated and are
parameterized by the LEC ${\tilde C}$. As noted above,
this constant includes the effects of
non-logarithmically enhanced two-loop contributions,
heavy mesons such as the $\eta$ and $\eta^\prime$ which have been
integrated out, and other
non-perturbative dynamics. On general grounds,
one could expect its natural size to be of order unity. A
comparison of Eq.(\ref{eq:chiral}) with the results of model calculations
is consistent with this expectation. For example, the model calculation of
Ref. \cite{bij01} corresponds roughly to ${\tilde C}\simeq 1$. Rigorously
speaking,
however, the precise value -- as well
as the sign -- of ${\tilde C}$ is unknown. An uncertainty $\Delta{\tilde
C}=\pm 1$ corresponds to
$\Delta \amulbl = \pm 31 \times 10^{-11}$,
which is roughly  one fifth of a standard deviation for the published
Brookhaven measurement\cite{brown}. One should not, however, treat this as
an estimate
of the theoretical
uncertainty in $\amulbl$. A value of ${\tilde C}$ equal to $+3$ or $-3$,
for example, would not be unusual.

Alternatively, one may use the experimental result for
$a_\mu$ to determine ${\tilde C}$.  To that end, we use the up-dated
results for hadronic
vacuum polarization
contributions
\cite{narrison}, the QED and electroweak loop contributions in Ref.
\cite{wjm} and the value for $\amulbl_{\rm chiral}$ given in Eq.
(\ref{eq:chiral}).
>From the E821 result for $a_\mu$ we obtain
\begin{equation}
\label{eq:c}
{\tilde C} = 7 \pm 5 \pm 3\ \pm 2\ \ \ ,
\end{equation}
where the first uncertainty arises from the experimental error in $a_\mu$,
the second corresponds to
the theoretical QED, electroweak, and hadronic vacuum polarization errors,
and the final uncertainty arises from the error in $\chi$. In the future,
the first uncertainty will be considerably
reduced upon complete analysis of the full E821 data set.
The value of
${\tilde C}$ is consistent with unity, though it could be considerably
larger, given the
other experimental and theoretical inputs into Eq. (\ref{eq:c}). Using the
second
solution for $\chi$ and $\amulbl$ gives ${\tilde C}=16\pm 5 \pm 3\ \pm 2$.

At present there is no indication that the hadronic
LEC ${\tilde C}$ differs substantially from its natural
size and, thus, no reason to discern effects of new physics, such as
loops containing supersymmetric particles\cite{wjm}, from
the $a_\mu$ result. In principle, a systematic calculation of some
of the effects arising at higher
order  -- such as terms of ${\cal O}(\alpha^3 N_C^0 p^2/\Lambda^2)$
enhanced by large logarithms
-- could modify this conclusion \cite{deraf2}. Similarly,
should the full E821 data imply a value for ${\tilde
C}$ which differs
significantly from $\pm 1$ ({\em e.g.}, by an order of
magnitude), one might argue
that there is evidence of new physics. Such a conclusion would
presumably require considerable
disagreement between the published E821 result\cite{brown} and the analysis
of the full
data set.
The most convincing analysis, however, would rely on a first
principles QCD calculation of $\amulbl$, a prospect which seems to lie well
into the future.

\bigskip
We thank  M. Knecht, A. Nyffeler,  M. Perrottet, and E. de Rafael, for
pointing out
terms omitted in a previous version of the manuscript. We also thank
D. Hertzog,  A. Kurylov,  and W. Marciano for useful discussions. This
work was supported in part by the U.S. Department of Energy contracts No.
DE-FG02-00ER41146 and DE-FG03-92ER40701 and National Science Foundation
Grant No. PHY-0071856.

\bigskip




\begin{figure}
\begin{center}
\includegraphics[height=3.5in]{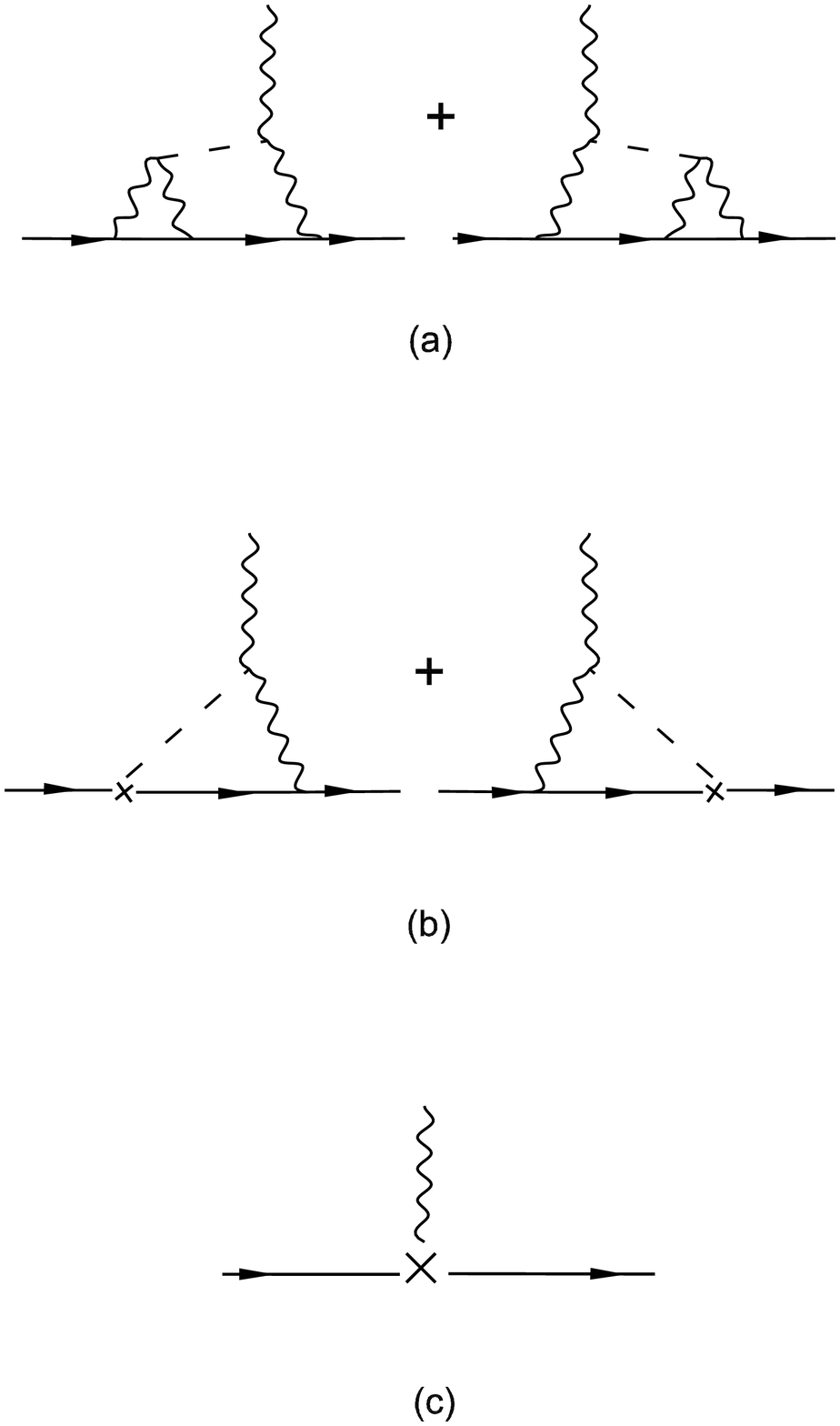}
\end{center}
\caption{\label{Fig1} Hadronic light-by-light contributions to muon
anomalous magnetic moment,
$\amulbl$, obtained from insertion of the WZW interaction at the
$\pi^0\gamma\gamma$ vertices. The $\times$ in (b) indicates the insertion
of the the
low-energy constant $\chi$ plus its counterterm. The $\times$ in (c)
denotes the magnetic moment coupling
 $C$ plus its counter term. The solid, dashed, and wavey lines denote the
$\mu$, $\pi^0$, and $\gamma$,
respectively.}
\end{figure}

\begin{figure}
\begin{center}
\includegraphics[height=3.5in]{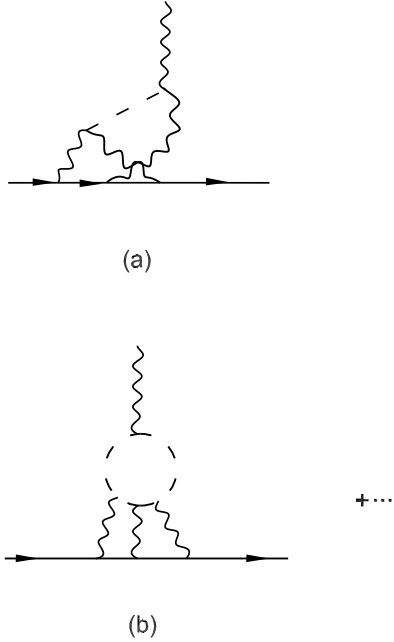}
\end{center}
\caption{\label{Fig2} Some hadronic light-by-light contributions to muon
anomalous magnetic moment that are
not enhanced by large logarithms.}
\end{figure}

\end{document}